\documentclass[runningheads]{waica}
\usepackage[T1]{fontenc}
\usepackage{graphicx}
\usepackage{hyperref}
\usepackage{color}

\usepackage{booktabs} 
\usepackage{array}    
\usepackage{tabularx}
\usepackage{caption} 
\usepackage{enumitem}
\begin{document}
\title{AI Supply Chain Galaxy: 3D Visual Analytics for License Compliance}
%

%
%

\author{Weiru Han\inst{1} \and
Xuetao Shi\inst{1} \and
Wenyi He\inst{1} \and
Wei Wang\inst{1} \and
Rui Zhao\inst{2} \and
Moming~Duan\inst{1}\thanks{Corresponding author: mmduan@dase.ecnu.edu.cn.}
}

\authorrunning{Weiru Han, Xuetao Shi et al.}

\institute{East China Normal University, Shanghai 200062, China
\and
Tianjin University, Tianjin, China}

\maketitle            
\begin{abstract}
The rapid proliferation of machine learning model reuse has transformed the AI ecosystem into a highly interconnected supply chain. Traditional compliance tools and static reports struggle to navigate these massive, multi-hop dependency networks. To address this, we present AI Supply Chain Galaxy (AISCG), an interactive 3D visual analytics system for model provenance and compliance auditing. AISCG maps models into a 3D spatial layout, integrating explicit structural dependencies with a rule-based compliance engine. It supports multi-scale exploration, from global community detection to localized, path-aware lineage tracing. We demonstrate its efficacy through an ecosystem-scale empirical analysis of 908,449 models from Hugging Face. Our findings reveal a concerning landscape: 55.46\% of models exhibit compliance risks or metadata conflicts/omissions. We also identified distinct risk patterns, including a 56.67\% license omission rate in adapter derivations and an 8.05\% ``license drift'' rate in fine-tuning. Through a case study on the complex Llama model family, we show how AISCG empowers analysts to intuitively trace inherited restrictive terms and identify root causes across deep topological networks, significantly reducing the cognitive load of compliance auditing.

\end{abstract}
\section{Introduction}
In recent years, the production, release, reuse, and creation of derivative models from machine learning (ML) models have entered a stage of large-scale proliferation. Model platforms, such as Hugging Face, have amassed millions of model assets. Through methods such as fine-tuning, adapters, quantization, and merging, these components have formed a highly interconnected reuse network~\cite{ref2,ref3}. This development paradigm centered on model reuse significantly lowers the barrier to AI application. However, it simultaneously complicates model provenance tracking and compliance auditing~\cite{ref4}. In scenarios involving multiple model and data sources, the AI supply chain has evolved from linear dependencies into a complex network structure~\cite{ref1,ref3}.

The expansion of technical reuse is accompanied by a corresponding increase in the complexity of license governance. The current ML ecosystem is characterized by the concurrent application of traditional Open Source Software (OSS) licenses (e.g., Apache-2.0, MIT, GPL), content licenses (e.g., Creative Commons Licenses), and AI model-specific licenses (e.g., OpenRAIL, Llama license family)~\cite{ref4,ref5,ref10}. However, these licenses exhibit significant disparities regarding applicable objects, definitions of derivatives, redistribution obligations, use restrictions, and liability terms. Existing studies indicate that issues such as license mismatch, license proliferation, and license conflicts are prevalent in the ML domain~\cite{ref4,ref5}. A notable instance is ``license drift'', where the licensing terms of a derivative model diverge from its upstream ancestors---for example, through the replacement of restrictive obligations with permissive ones. It becomes difficult to determine where legal risks originate when models depend on multiple types of derivatives. This complexity allows compliance risks to silently propagate downstream along deep dependency chains. Beyond explicit links, the AI supply chain is further complicated by implicit dependencies such as knowledge distillation and data-driven inheritance~\cite{ref4,ref5}. While identifying these remains an open research question, providing a transparent view of the explicit reuse network is a critical and foundational first step for compliance auditing. Traditional Software Composition Analysis (SCA) tools (e.g., Black Duck, FOSSology)~\cite{ref18,ref20}, which primarily rely on code-level dependency linking or static signature scanning, are inadequate for handling such a massive and entangled ecosystem~\cite{ref1,ref4}.

Although existing studies have examined the model ecosystem and its compliance challenges~\cite{ref1,ref3,ref4,ref10}, there is a lack of systematic support for integrated and interactive compliance analysis. On one hand, supply chain research primarily provides macro statistics and identifies general trends in the ecosystem~\cite{ref1,ref2,ref3}. On the other hand, license studies focus on the extraction of rules and the assessment of legal compatibility~\cite{ref4,ref5,ref10}. These two areas remain largely separate, as legal frameworks have not yet been integrated with the large-scale dependency data found in model repositories.
For auditors, tracing a specific legal risk through a long chain of models requires the manual correlation of textual rules with structural data. This process is difficult to perform using only static reports or offline summaries. Therefore, an integrated approach is needed to help users navigate the relationship between legal constraints and model lineages, which leads to the requirement for more effective methods of data exploration.

Visual analytics is essential for transforming abstract metadata into actionable license compliance insights across multiple scales. License compliance states are dynamic properties that propagate through model lineages~\cite{ref4}. At the local lineage level, visualizing these specific paths allows auditors to intuitively trace restrictive obligations and pinpoint exact conflict origins. At a global systemic level, mapping the structural modularity of the AI supply chain reveals dense model communities and helps analysts identify core reuse hubs. Therefore, an effective visual analytics approach for license compliance should bridge these scales, coupling global ecosystem structures with localized, path-aware compliance details.

\begin{figure}[t]
\centering
\includegraphics[width=\textwidth]{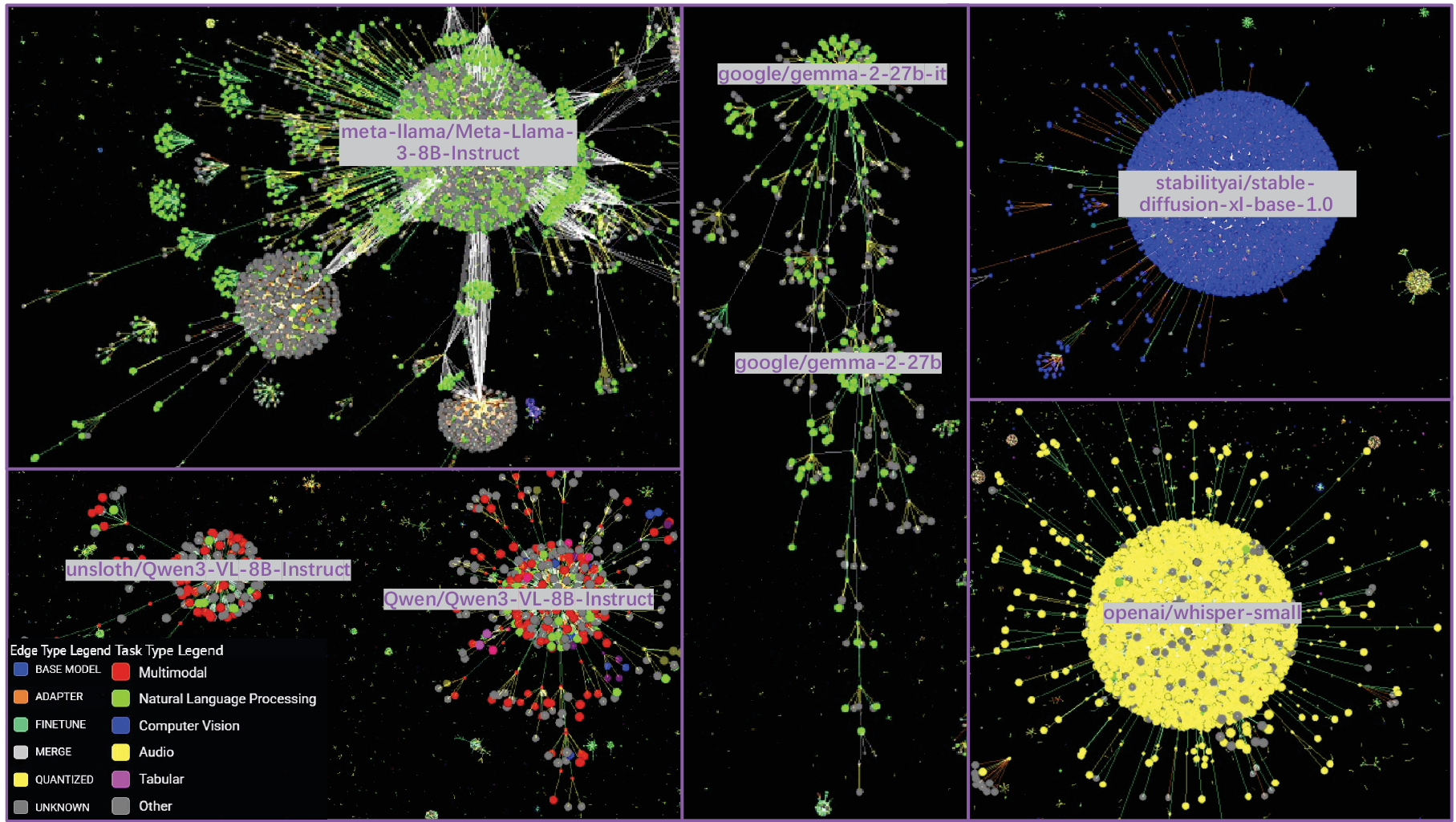}
\caption{AI supply chain visualized in AISCG. Node colors encode application domains, while edge colors represent distinct reuse methods.}\label{fig6}
\end{figure}

In this paper, we propose AI Supply Chain Galaxy (AISCG), a 3D visual analytics system for model provenance and license compliance. The system maps millions of models into a 3D spatial layout where node geometry (size and color) encodes centrality and risk states, while orbital edges represent diverse reuse dependencies (Fig.~\ref{fig6}). We provide three core capabilities: (1) Structural Insight, revealing model communities through spatial expansion; (2) License compliance analysis, providing a license compliance report for each model; and (3) Lineage Tracing, isolating multi-order lineage paths to pinpoint legal origins. Supporting both desktop and mobile navigation, our system offers an intuitive environment that facilitates interactive compliance exploration for auditing compliance in complex AI supply chains. Compared with static reports and table-based summaries, our system integrates structural exploration, compliance status tracking, and interactive compliance analysis in a unified workflow. This provides a more intuitive entry point for compliance analysis in complex model-reusing scenarios.

The main contributions of this paper are summarized as follows:

\begin{itemize}[label=$\bullet$, topsep=0pt]
    \item \textbf{A Multi-Dimensional Framework:} We propose a novel visual analytics framework that integrates model dependencies with legal compliance logic across spatial, temporal, and topological dimensions.

    \item \textbf{The AI Supply Chain Galaxy System:} We implement a 3D visual analytics system providing three core capabilities: community detection, license compliance analysis and provenance recovery.

    \item \textbf{Large-Scale Empirical Insights:} Analyzing 908,449 models, we find that 55.46\% exhibit compliance risks. We identify distinct risk patterns, such as a 56.67\% license omission rate for child models in adapter-based relationships and an 8.05\% ``license drift'' rate in fine-tuning edges.
\end{itemize}





\section{Related Work}
\subsection{AI Model Ecosystem and Supply Chain Analysis}
Recent studies consistently show that the ML ecosystem has evolved from isolated model releases into a large-scale reuse network. Jiang et al. first characterized the pre-trained model supply chain across multiple hubs and highlighted that model ecosystems already exhibit software-supply-chain-like risk surfaces~\cite{ref1}. Their follow-up study further provided empirical evidence of reuse workflows on Hugging Face, including provenance, reproducibility, and trust-related challenges in real-world model reuse~\cite{ref2}. At larger scale, Laufer et al. analyzed nearly two million Hugging Face models and revealed rich family-tree structures formed by fine-tuning, adapters, quantization, and merging~\cite{ref3}. 

From a supply-chain risk perspective, prior security work shows reuse and outsourcing can introduce persistent downstream vulnerabilities. BadNets showed that backdoors injected during outsourced or transfer-based workflows can remain effective even after downstream adaptation~\cite{ref22}. This persistent vulnerability demonstrates that model lineage can act as a risk-propagation channel. 



A parallel line of work emphasizes model lineage and ecosystem cartography. Horwitz et al. proposed unsupervised model tree heritage recovery directly from weights, showing that lineage can be inferred even when metadata is incomplete~\cite{ref21}. Their Model Atlas position further argues for a population-level graph representation of models, attributes, and transformations, enabling forensics and ecosystem-scale analysis~\cite{ref23}. 
Complementing this, Wu et al.~\cite{wu2026llm} derive ``LLM DNA'' from behavioral signatures to identify undocumented lineages across architectures without requiring weight access.
This direction is consistent with broader calls for model transparency and documentation (e.g., model cards)~\cite{ref9}, but also highlights a key limitation of current practice: documentation is often missing, sparse, or inconsistent at scale~\cite{ref2,ref6,ref21,ref23}.

Overall, existing research has established three facts: (i) model reuse is massive and structurally complex~\cite{ref1,ref2,ref3}, (ii) supply-chain risks can propagate along reuse paths~\cite{ref1,ref22}, and (iii) provenance signals are frequently incomplete, motivating lineage recovery via weight-based~\cite{ref21} or behavioral~\cite{wu2026llm} fingerprinting and atlas-style modeling~\cite{ref23}. However, most prior work remains measurement-oriented or method-oriented, with limited support for interactive, path-level analysis that jointly links topology, provenance, and attributes related to compliance. This gap motivates our visual analytics approach for AI model supply chain exploration and auditing.

\subsection{License Compliance in Machine Learning}
Researchers increasingly recognize license compliance in ML as a distinct problem that existing OSS compliance pipelines cannot fully address. In contrast to software-only projects, ML systems typically combine heterogeneous assets (code, datasets, models, and generated artifacts) under different licensing regimes. These regimes encompass a complex mix of OSS licenses, content licenses, and model-specific licenses~\cite{ref4,ref5,ref10,ref19}. As model reuse scales through fine-tuning, adapters, quantization, and merging, legal obligations propagate across multi-step dependency chains. Consequently, rights and obligations become significantly harder to interpret compared to conventional file-linking scenarios~\cite{ref4,ref5}.

Recent work has begun to formalize this problem space. ModelGo presents one of the first practical analyses tailored to ML workflows and reports real-world conflict patterns in mixed-license reuse settings~\cite{ref4}. Building on this line, subsequent studies further argue that current model licensing practice suffers from systematic issues, especially license mismatch, proliferation, conflict. They provide evidence that widely used OSS or content licenses are often semantically under-specified for model publishing and derivative governance~\cite{ref5,ref19}. In parallel, behavioral-use licensing (e.g., RAIL-style restrictions) expands the compliance landscape beyond classical redistribution obligations, introducing enforceable use-constraint terms that are uncommon in traditional OSS ecosystems~\cite{ref10}. These developments indicate that ML compliance is not only a compatibility-checking problem, but also a workflow- and policy-aware reasoning problem.

Prior OSS literature remains important as methodological foundation. SPDX-based compatibility automation and large-scale conflict studies in software ecosystems provide transferable ideas for rule encoding, incompatibility graphs, and conflict detection~\cite{ref18,ref20}. However, their assumptions do not always hold in ML settings: model artifacts often lack explicit embedded license texts; dependencies can be implicit (e.g., representation transfer rather than direct code inclusion); and obligations may involve interactions among model, data, and downstream outputs~\cite{ref4,ref5,ref19}. Therefore, OSS techniques are necessary but insufficient as end-to-end solutions for ML compliance.

A further emerging frontier concerns generative systems and AI-assisted development. LiCoEval shows that LLMs can produce code with similarity to licensed sources while failing to provide adequate license information, revealing a new operational compliance gap at generation time~\cite{ref11}. At the ecosystem governance level, policy analyses also highlight persistent ambiguity in how “open” is interpreted in AI practice, which complicates compliance expectations and enforcement boundaries~\cite{ref7}. Together, these findings reinforce the need for compliance analysis approaches that are both technically grounded and context-aware.

Overall, existing work has established the urgency of ML license compliance and identified key risk patterns, but there is still limited support for analyst-facing, interactive environments that connect legal rules with concrete multi-hop dependency structures. This gap motivates our approach, which couples compliance reasoning with explicit model provenance and dependency exploration in a unified analysis workflow.

\subsection{Visual Analytics for Provenance and Dependency Networks}
Recent empirical studies have documented the scale and structural complexity of the machine learning ecosystem. Laufer et al.~\cite{ref3} analyzed nearly two million models on Hugging Face and identified rich family-tree structures formed through fine-tuning, adapters, quantization, and model merging. Horwitz et al.~\cite{ref21} explored model heritage recovery when metadata is incomplete, and their later “Model Atlas” perspective~\cite{ref23} called for a population-level graph representation to support ecosystem-scale lineage tracking.

Visualizing such large, dense, and heterogeneous graphs remains a central challenge in graph analytics. The survey by von Landesberger et al.~\cite{ref17} shows that high-density topologies can exceed the practical readability of conventional 2D node-link views. Prior perceptual and empirical work further suggests that interactive 3D environments may provide task-dependent benefits, particularly for path-tracing and cluster-identification tasks in complex networks~\cite{ref26,ref27}. 

For machine learning analysis interfaces, Bäuerle et al.~\cite{ref29} emphasize the role of Multiple Coordinated Views (MCVs) in supporting workflow-oriented exploration and inspection. In the context of provenance and dependency networks, this perspective indicates that no single view is sufficient for both global topology understanding and local path-level examination. Following this design rationale, we combine a spatial graph view with coordinated detail views to support multi-scale exploration of model lineages and dependency relations. 

\section{Design}

\subsection{Architecture and Data Pipeline}
\begin{figure}[t]
\centering
\includegraphics[width=\textwidth]{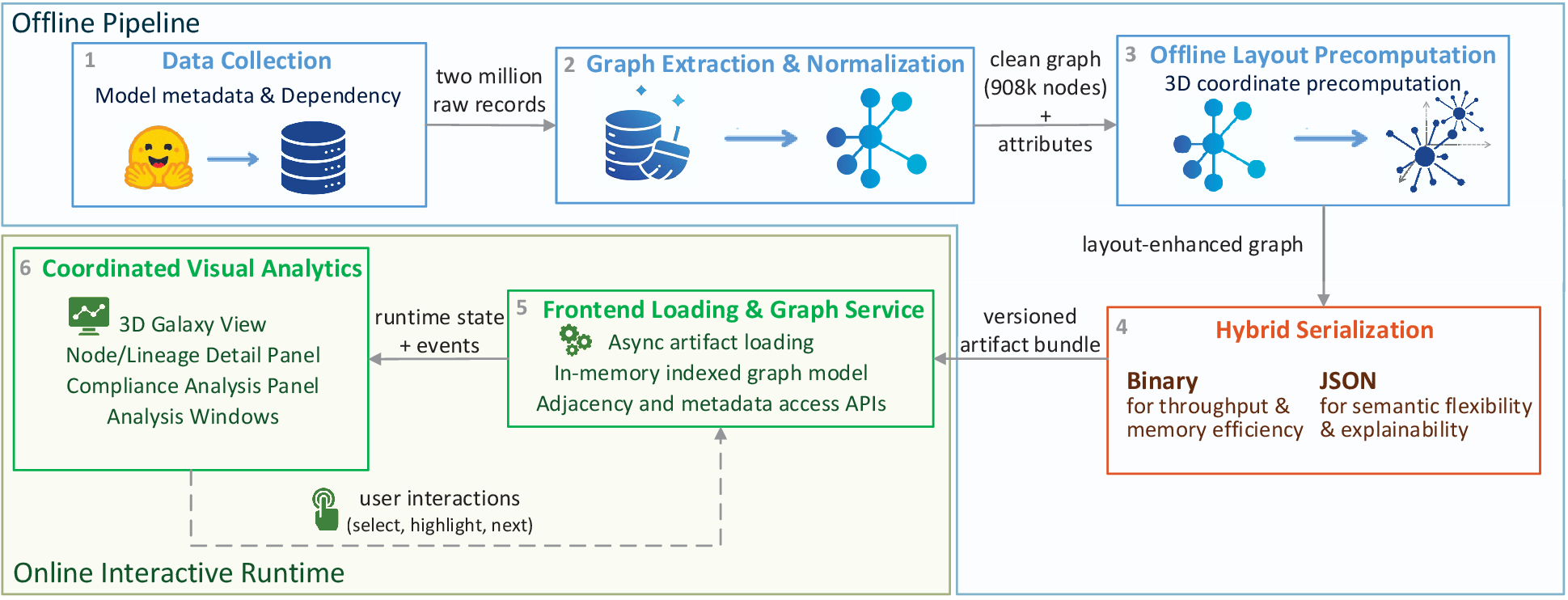}
\caption{Architecture and Data Pipeline of AI Supply Chain Galaxy.}\label{fig1}
\end{figure}

AI Supply Chain Galaxy follows an offline-online architecture. Offline processing is responsible for data collection and graph extraction, while online runtime focuses on interactive visual analytics (Architecture see
Fig.~\ref{fig1}).

Offline stage. The pipeline ingests pre-collected model metadata and dependency relations, normalizes heterogeneous license fields into project-level canonical keys, builds a directed reuse graph, precomputes 3D node coordinates via an offline force-directed layout, then serializes graph artifacts into deployment-ready files: binary arrays for high-volume structural/spatial data, JSON objects for semantic attributes and compliance outputs.

Online stage. The frontend asynchronously loads these artifacts, reconstructs in-memory graph views, serves low-latency interaction primitives: neighborhood query, shortest-path tracing, node-centric compliance inspection. An event-driven runtime syncs rendering and analytical panels, so node selection, highlighting, windowed diagnostics update consistently in a unified workflow.

We adopt a hybrid binary/JSON storage strategy (Table.~\ref{tab1}). This design separates dense numeric arrays (binary for speed and memory efficiency) from semantically rich attributes (JSON for flexibility and interpretability), enabling fast loading without sacrificing explainability.

\begin{table}[h]
\centering
\caption{Hybrid data storage strategy for scalable graph artifacts.}\label{tab1}
\renewcommand{\arraystretch}{1} 
\scriptsize %
\begin{tabular}{@{} l >{\raggedright\arraybackslash}p{8.5cm} @{}} 
\toprule
\textbf{Artifact Name} & \textbf{Description and Semantic Content} \\
\midrule
\addlinespace[2pt]
\texttt{positions.bin}  & Precomputed 3D coordinates (dense numeric arrays). \\
\texttt{links.bin}      & Compact adjacency encoding for high-throughput parsing. \\
\texttt{link\_data.bin} & Flattened edge attribute arrays (source/target/type IDs). \\
\addlinespace[2pt]
\texttt{labels.json}    & Node labels used for search and interactive display. \\
\texttt{nodeData.json}  & Semantic metadata (license, author, tags, timestamps). \\
\texttt{link\_types.json} & Dependency relation type dictionary. \\
\texttt{manifest.json}  & Version indexing and artifact mapping entry. \\
\bottomrule
\end{tabular}
\end{table}

\subsection{Compliance Analysis Rules}
To operationalize license compliance over AI model reuse graphs, we implement an offline rule-based analysis module. After normalizing raw license metadata into canonical license keys, the engine evaluates compliance risks across predefined propagation edge types (Finetune, Merge, Quantize, Adapter) and generates node-level risk records. The current implementation applies eight rules under three analytical dimensions (Table.~\ref{tab3}): Local Integrity captures node-level metadata defects and semantic mismatches; Propagation Consistency models the transfer of obligations and restrictions via downstream BFS traversal; and Provenance Legality verifies whether derivative license claims are supported by official upstream lineage. Each triggered rule produces structured, explainable risk entries with assigned severity levels (Warning/Error), enabling interpretable root-cause tracing for complex compliance failures. This module is designed to provide heuristic risk indications rather than formal legal adjudication.

\begin{table}[h]
\centering
\caption{Compliance analysis rules mapped across three analytical dimensions.}\label{tab3}
\renewcommand{\arraystretch}{1}
\setlength{\tabcolsep}{6pt}
\scriptsize
\begin{tabularx}{\textwidth}{@{} l l >{\raggedright\arraybackslash}X @{}}
\toprule
\textbf{Analytical Dimension} & \textbf{Rule} & \textbf{Detection Goal and Risk Description} \\
\midrule
\textbf{Local Integrity} 
    & \texttt{Mismatch} 
    & Detects missing, unknown, or semantically inconsistent metadata (e.g., naming cues conflicting with declared tags). \\
\addlinespace
\textbf{Propagation Consistency} 
    & \texttt{Copyleft\_Terms} 
    & Identifies the loss or relabeling of restrictive terms during derivation. \\
    & \texttt{Copyleft} 
    & Detects inconsistencies in strict copyleft license propagation. \\
    & \texttt{Conflict\_ND} 
    & Flags potential derivative-use conflicts under ND-constrained upstream sources. \\
    & \texttt{Conflict\_CC} 
    & Identifies predefined incompatibility patterns between source and target CC licenses. \\
\addlinespace
\textbf{Provenance Legality} 
    & \texttt{Conflict\_FSF} 
    & Verifies downstream claims against GPL-family upstream constraints. \\
    & \texttt{Conflict\_La2E}/\texttt{La3E} 
    & Checks whether Llama-2/3 license claims are supported by official lineage ancestry. \\
\bottomrule
\end{tabularx}
\end{table}

\subsection{Visual Encoding and Layout}

\begin{figure}[t]
\centering
\includegraphics[width=\textwidth]{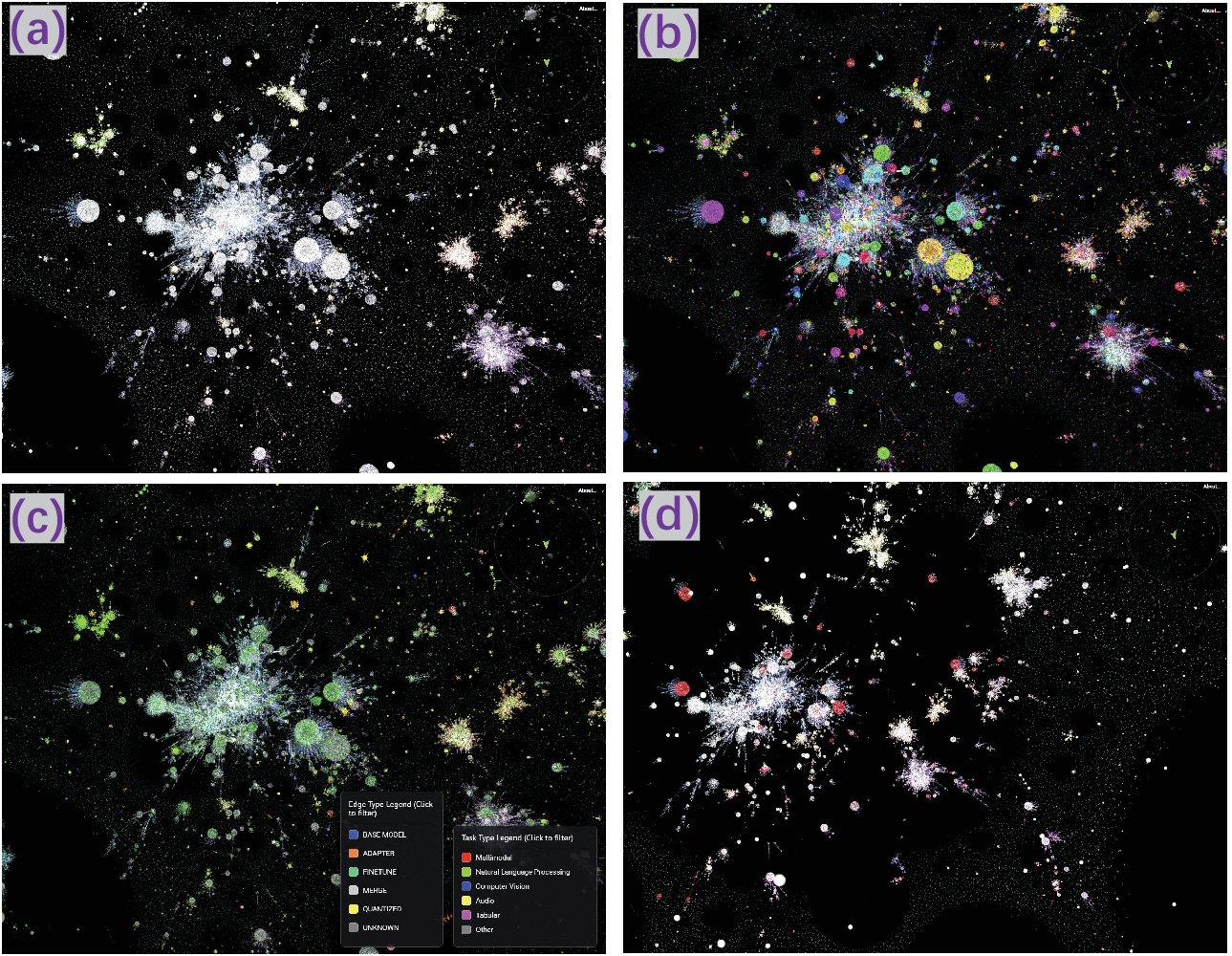}
\caption{Multi-faceted Visual Analysis Modes in AISCG, (a) Default View with edge-based reuse types, (b) Community View for structural clustering, (c) Task Type View with interactive filtering, and (d) Risk View highlighting license compliance issues.}\label{fig4}
\end{figure}

To facilitate the exploration of massive and entangled model supply chains, we design a multi-modal visual encoding scheme. By mapping different semantic attributes to visual channels, our system provides four analysis views (Fig.~\ref{fig4}), enabling analysts to switch perspectives across structural, ecological, and risk dimensions.

The system adopts an offline precomputed 3D graph layout. In the Default View (Fig.~\ref{fig4}a), we establish the structural baseline:

Node Encoding: All nodes are rendered in white to maintain visual neutrality, focusing the user's attention on the network topology. Node size encodes structural importance (e.g., in-degree or centrality proxy), highlighting the ``hubs'' of the supply chain.

Edge Encoding: Color represents the reuse relationship type (e.g., Fine-tuning, Merging, Adapter). This allows analysts to perceive the technical patterns of model evolution at a glance.

To uncover the macro-level organization of the AI ecosystem, we provide three categorical views:

(1) Community View (Fig.~\ref{fig4}b) applies color mapping based on structural clustering. This facilitates the rapid identification of community and cross-community bridges.
(2) In Task Type View (Fig.~\ref{fig4}c), nodes are color-coded by their application domain (e.g., NLP, Multimodal, Computer Vision). This view features an interactive legend (bottom-right of Fig.~\ref{fig4}c). Users can dynamically toggle specific node categories or edge types. This capability can help reduce visual clutter and focusing on domain-specific supply chains (e.g., the Merging paths within the NLP domain).
(3) The Risk View (Fig.~\ref{fig4}d) transforms legal risks into immediate visual signals. While maintaining the topological context, the system renders models with detected license conflicts in alarm red. Through this view, analysts can perceive the spatial distribution of compliance risks.

\subsection{Interactive Compliance Workflow}

\begin{figure}
\centering
\includegraphics[width=\textwidth]{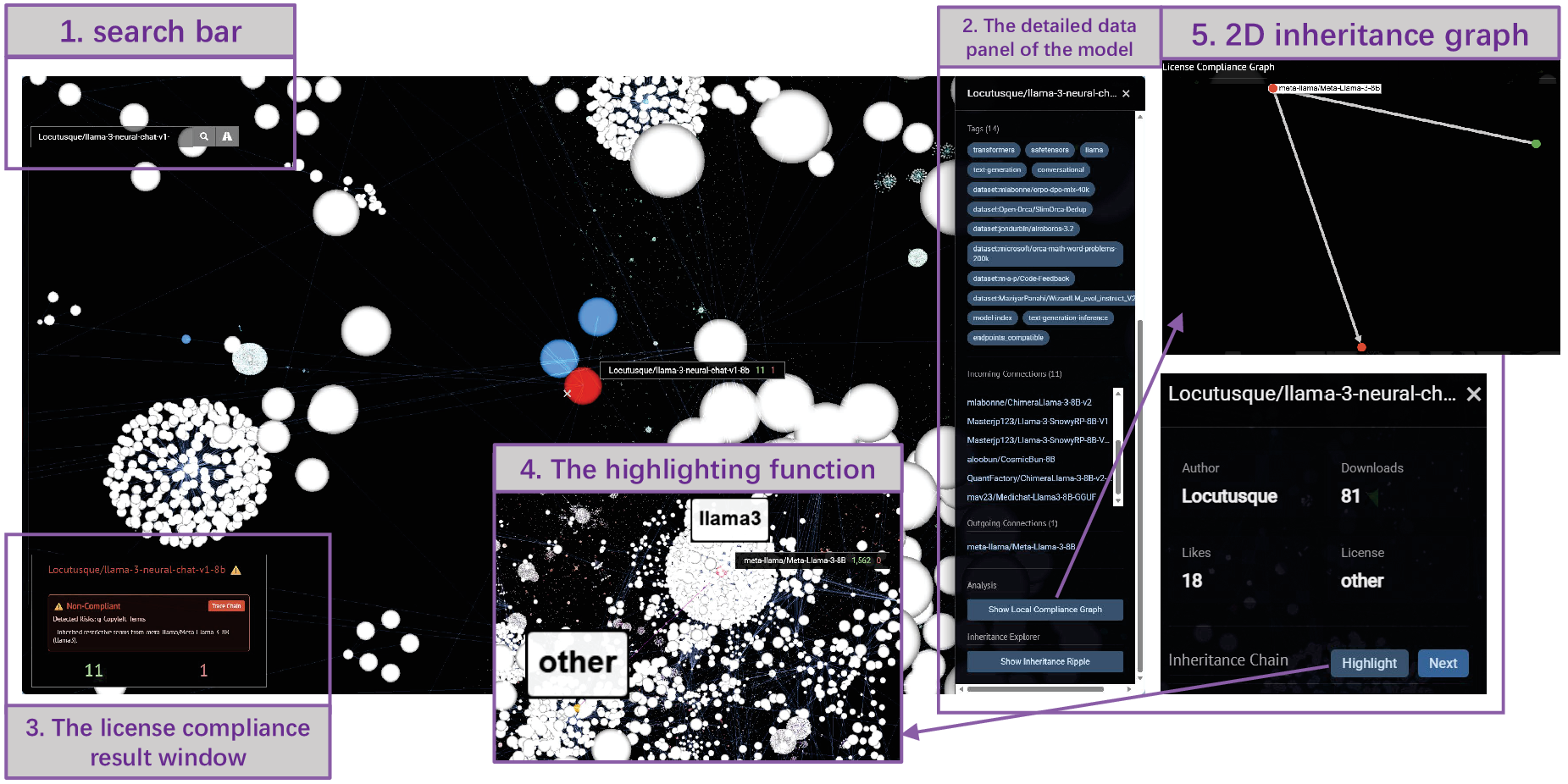}
\caption{The analysis Interface of AISCG, featuring coordinated views for searching, auditing, and lineage tracing.}\label{fig5}
\end{figure}
We organize interactions around a workflow from global exploration to localized compliance decisions. (Fig.~\ref{fig5})
Users start from the global 3D galaxy view, utilizing pan/zoom/rotate navigation to discover macro-structures such as hubs and dense communities. To locate a specific target, the system supports focal point retrieval via direct node clicking or the Search Bar. Upon searching, the camera smoothly interpolates to the target's coordinates.

After selecting a focal node, the system provides a multi-faceted compliance report. The Detailed Data Panel displays model metadata (e.g., downloads, likes, license tags) and local connectivity. If the compliance engine detects a violation, the License Compliance Result Window explicitly lists the triggered risk rules (e.g., \texttt{Copyleft\_Terms}). Crucially, the system identifies the upstream origin of the risk. For instance, as shown in Fig.~\ref{fig5}, the system reports: ``Inherited restrictive terms from meta-llama/Meta-Llama-3-8B (llama3)'', This attribution reduces the time required to pinpoint the root cause in deep supply chains.

To verify the diagnosed risks, the system provides two visualization aids for path tracing. The Highlighting Function illuminates the entire inheritance chain in the 3D space, showing the path from the root model to the current node. Simultaneously, a localized 2D Inheritance Graph flattens the genealogy, providing a schematic view of the multi-hop relationships. Through the ``Next'' and ``Highlight'' controls, analysts can perform a step-by-step traversal along the provenance path, manually inspecting the legal transitions at each generation.

\subsection{Design Discussion}
Our design follows three practical choices. First, we use 3D spatialization as the primary structural view to support macro-level exploration. Second, we adopt an offline-layout and online-interaction pipeline to avoid expensive runtime global recomputation and to keep exploration responsive at large scale. Third, we use event-driven modular coordination to synchronize rendering, lineage inspection, and compliance panels while preserving implementation extensibility.

The current system focuses on explicit, metadata-declared dependencies and rule-based compatibility checks. Therefore, the output should be interpreted as analyst support rather than formal legal adjudication. Implicit dependencies (e.g., distillation- or data-mediated inheritance) and broader legal interpretation remain out of scope and are left for future work.

\section{Empirical Analysis}
To evaluate the effectiveness of our visual analytics approach and to uncover the current state of model licensing practices, we conducted an ecosystem-scale empirical study. We applied our offline rule-based compliance engine to a massive dataset collected from the Hugging Face hub.

\subsection{Ecosystem-Scale Compliance Landscape}

The dataset comprises 908,449 non-independent models (i.e., models possessing explicit upstream or downstream derivation relationships). By executing the compliance rules defined in Section 3.2, we obtained a macro-level diagnostic of the AI supply chain.

Our evaluation reveals a concerning ecosystem: 55.46\% of the analyzed models exhibit at least one compliance risk or license annotation conflict.
As summarized in Table.~\ref{tab2}, the vast majority of compliance failures stem from local metadata quality issues. The Local Integrity rule (Mismatch) flagged 459,745 models, accounting for 50.6\% of the total dataset. This indicates a pervasive phenomenon where derivative models are published with missing or unknown licensing information.

\begin{figure}[t]
\centering
\includegraphics[width=0.9\textwidth]{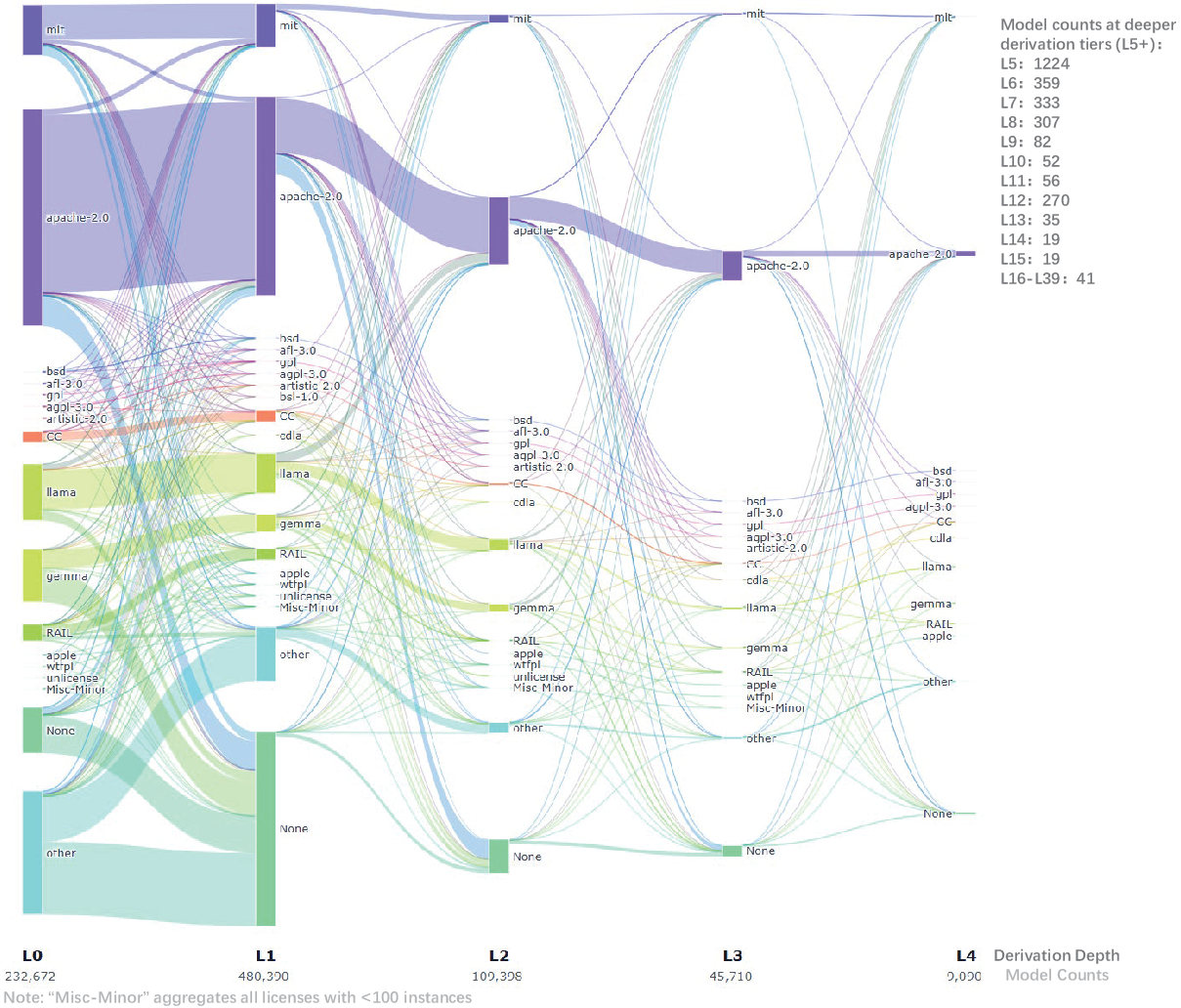}
\caption{License evolution across successive generations in the model supply chain. The Sankey diagram visualizes the propagation and transformation of licenses from root models (L0) to their derivatives across five tiers (L0-L4). The horizontal axis represents the derivation depth, while vertical blocks represent license categories. Flows between tiers indicate license inheritance or changes; the width of each flow is proportional to the model count.} \label{fig2}
\end{figure}

Furthermore, downstream obligation failures are common. Approximately 13.2\% (119,898 models) triggered the Restrictive Terms Continuity rule (Copyleft\_Terms). Downstream models fail to inherit mandatory restrictive clauses from their upstream ancestors (e.g., a derivative explicitly based on Llama 3 being incorrectly relabeled as Apache-2.0). 

To visualize how compliance risks propagate across generations, we extracted a ``forest'' comprising 880,057 models, excluding multi-parent merges to ensure a clear, single-lineage evolutionary path. As illustrated in Fig.~\ref{fig2}, the Sankey diagram depicts the inter-generational dynamics of license inheritance across five tiers (L0 to L4). 
The visualization results show that from L0 to L1, a large number of models lost their licenses during the first derivation. Such as the Gemma family and models categorized under the ``other'' license, half of their child models lose their license metadata during the first derivation. 50.6\% of the models have ``Mismatch'' problem, many of which are caused by the loss of license information during inter-generational transfer. Even if the root node has an extremely permissive Apache-2.0 license, there is still traffic going into ``None'' (cyan strip). At the same time, we can also clearly see that for models with licenses containing restrictive terms, their derived models have changed the licenses to more permissive ones such as Apache-2.0 and MIT. This phenomenon can be observed in all four derivations in the figure.


\begin{figure}[ht]
  \begin{minipage}[b]{0.42\linewidth}
    \centering
    \captionof{table}{Distribution of triggered compliance risks across the analyzed dataset.}\label{tab2}
    \renewcommand{\arraystretch}{0.9} 
    \scriptsize
    \begin{tabular}{@{} l r r @{}}
    \toprule
    \textbf{Risk Category} & \textbf{Models} & \textbf{\%} \\
    \midrule
    \texttt{Mismatch}        & 459,745 & 50.61 \\
    \texttt{Copyleft\_Terms} & 119,898 & 13.20 \\
    \texttt{Conflict\_La3E}  & 8,924   & 0.98  \\
    \texttt{Conflict\_La2E}  & 3,883   & 0.43  \\
    \texttt{Copyleft}        & 3,866   & 0.43  \\
    \texttt{Conflict\_ND}    & 3,378   & 0.37  \\
    \texttt{Conflict\_CC}    & 1,554   & 0.17  \\
    \texttt{Conflict\_FSF}   & 267     & 0.03  \\
    \bottomrule
    \end{tabular}
    \vspace{25pt} 
  \end{minipage}
  \hfill
  \begin{minipage}[b]{0.56\linewidth}
    \centering
    \includegraphics[width=\textwidth]{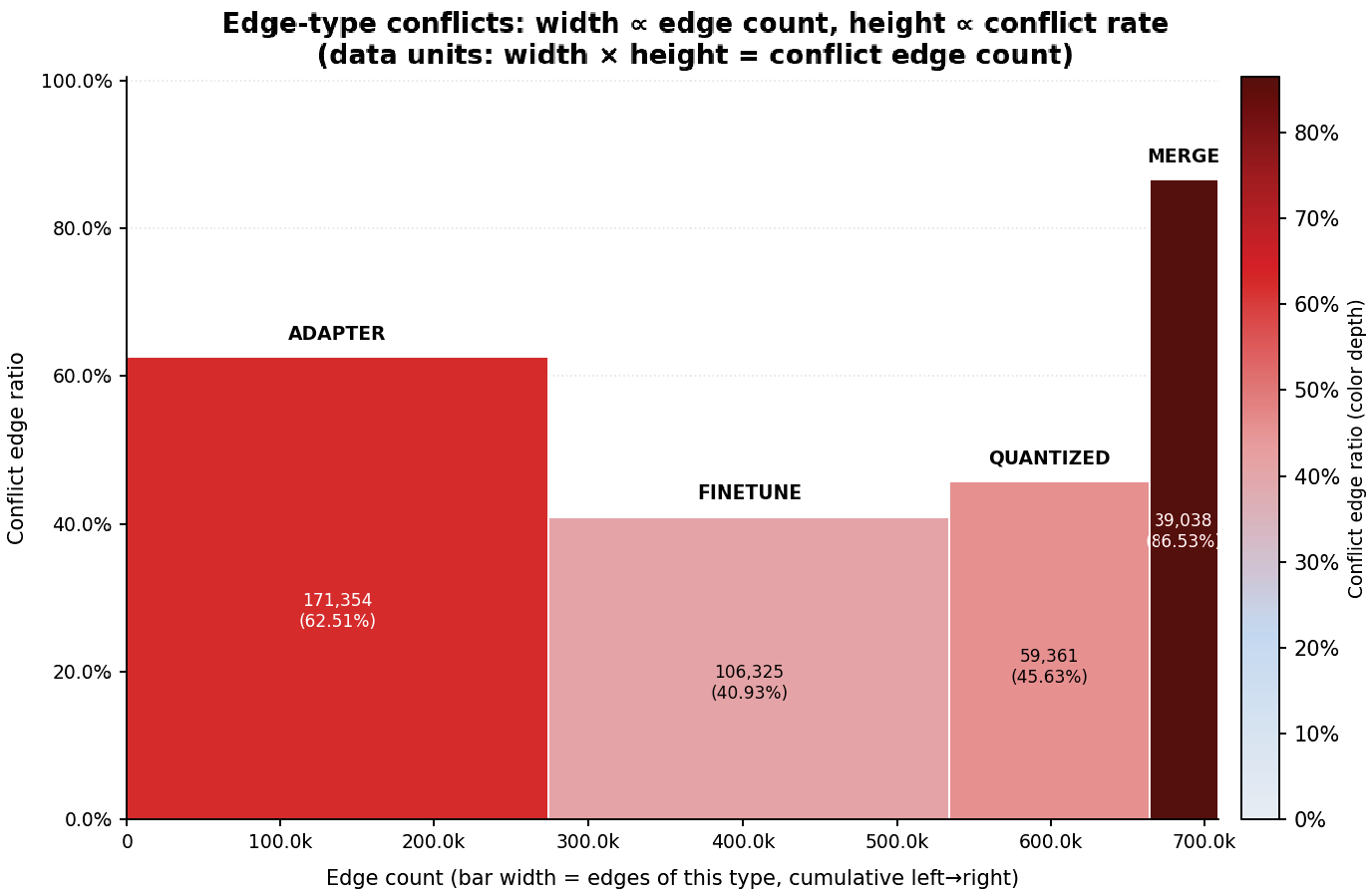}
    \caption{Conflict rates across reuse topologies.} \label{fig3}
  \end{minipage}
\end{figure}

\subsection{Compliance Status of Different Types of Edges}

To understand how different technical reuse paradigms impact legal compliance, we analyzed the conflict rates across different dependency edge types. An edge is classified as a ``conflict edge'' if at least one of its connecting nodes exhibits a compliance violation. As illustrated in Fig.~\ref{fig3}, the risk varies significantly depending on the topological relationship. To uncover the causes of these conflicts, we investigated the incidence of missing licenses and ``explicit license drift'' (where restrictive upstream clauses are overwritten by permissive downstream licenses) in different types of edges.

While license omission is prevalent across the ecosystem, the degree of information loss varies significantly by reuse paradigm. Adapters have the highest proportion among all relationships and they exhibit the most striking disparity: while only 4.49\% of parent models lack a license, this figure surges to 56.67\% in their children, representing the sharpest relative increase among all categories. Fine-tuning follows a similar trend, where the proportion of missing licenses grows from 10.22\% in parent models to 25.92\% in children. Model merging, which carries the highest overall conflict rate (86.53\%), also displays the most severe license omission. The unlicensed rate of involved parent models (35.86\%) nearly doubles to 62.50\% in children. Finally, although quantization is relatively less affected, the increase from 21.96\% to 29.73\% still indicates a non-negligible risk of metadata attrition during automated conversion.

While fine-tuning has the lowest overall conflict rate (40.93\%), it exhibits a distinctly different failure mode. Fine-tuning presents the highest rate of explicit ``license drift'' among all operations. In 8.05\% of fine-tuning edges, developers replace a restrictive upstream license with a permissive one (compared to 4.06\% for quantization, 3.03\% for merge, and 1.59\% for adapters). We hypothesize this is driven by a behavioral bias: developers who invest significant computational resources and private data into fine-tuning may falsely assume they have generated an independent artifact, incorrectly claiming permissive ownership (e.g., relabeling a Llama derivative as Apache-2.0).

These macroscopic statistics demonstrate that AI compliance risks are heterogeneous, oscillating between metadata loss (in Adapters, Quantization and Merge) and legal misinterpretation (in Fine-tuning). Such variance underscores the necessity for interactive, path-aware analytics tools like AISCG.

\subsection{Case Study}

To further validate the capabilities of the AISCG in handling real-world dependencies, we conducted a targeted case study on the Llama model family. As one of the most prominent open-weights lineages, Llama presents an ideal test for our system due to its massive scale and labyrinthine derivation patterns. 

Beyond single-root lineages, we identified 28,392 models (approximately 3.1\% of the total dataset) constructed from diverse ancestries. The Llama family exhibits a high participation rate in these multi-root lineages at 26.31\%. These multi-root merges represent a critical compliance spot: combining weights from models governed by different licenses elevates the risk of legal incompatibility. To demonstrate AISCG's capacity to audit these complex multi-root topologies, we highlight the specific case of the model Kukedlc/NeuralLLaMa-3-8b-DT-v0.1.

On the surface, this model's static metadata only shows three direct parent nodes. However, it is actually influenced by 19 distinct upstream ancestor models. When examining this node in the License Compliance Result Window, the compliance engine triggers multiple alerts, including \texttt{Copyleft\_Terms}, \texttt{Conflict\_CC}, and \texttt{Conflict\_La3E}. Crucially, the system provides an interpretable, multi-hop attribution list, pinpointing exactly where these conflicts originated among the 19 ancestors. For example, the system traces: (1) Inherited restrictive terms spanning multiple generations back to the original meta-llama/Meta-Llama-3-8B; (2) A severe CC-BY-NC-4.0 (Non-Commercial) restriction inherited from Undi95/Llama-3-Unholy-8B; (3) Copyleft obligations originating from elyn-dev/\-Llama-3-Soliloquy-8B-v2 (CC-BY-NC-SA-4.0).

Through AISCG, manual traversal of multi-generation warehouse branches and cross-checking of multiple licenses are avoided, and the license compliance risk brought by risks deep inheritance paths involving various reuse methods are clearly presented.


\section{Conclusion}
In this paper, we presented AISCG, an interactive 3D visual analytics environment designed to address provenance and license compliance challenges in modern AI ecosystems. By integrating a rule-based compliance engine with topological visualization, AISCG enables intuitive multi-hop risk tracing. Our ecosystem-scale analysis of over 900,000 models revealed a concerning landscape: 55.46\% of assets exhibit compliance risks, featuring distinct patterns like ``license drift''. While current efforts focus on explicit metadata dependencies, future work will explore implicit dependencies (e.g., knowledge distillation) and NLP-based automated license parsing to further advance transparent AI model governance.

\begin{credits}
\subsubsection{\ackname}
This research is sponsored by Shanghai Pujiang Programme (Award No. 25PJA029).

\end{credits}
%
%
%
%

\end{document}